%% file: main.tex
\documentclass{article}
\newif\ifArxivMode
\ArxivModetrue
\usepackage[accepted]{mlsys2025}
\makeatletter
\renewcommand{\Notice@String}{Preprint.}
\makeatother
\input{paper/packages}

\usepackage{times}
\input{paper/macros}
\input{paper/metadata}

\hypersetup{colorlinks=true,linkcolor=blue,citecolor=blue,urlcolor=blue,pdfauthor={},pdftitle={\PaperPDFTitle}}
\ifArxivMode\hypersetup{pdfauthor={\PaperAuthorName},pdfsubject={Preprint}}\fi
\mlsyssetsymbol{intern}{*}
\mlsyssetsymbol{corresponding}{\ensuremath{\dagger}}
\newcommand{\EmptyPaperEmail}{}
\ifdefined\isaccepted
  \ifx\PaperEmail\EmptyPaperEmail
    \makeatletter
    \gdef\mlsyscorrespondingauthor@text{\PaperCorrespondingAuthor}
    \makeatother
  \else
    \mlsyscorrespondingauthor{\PaperCorrespondingAuthor}{\PaperEmail}
  \fi
\else
  \mlsyscorrespondingauthor{\PaperCorrespondingAuthor}{\PaperEmail}
\fi
\mlsystitlerunning{\PaperRunningTitle}
\begin{document}\twocolumn[
\mlsystitle{\PaperTitle}
\begin{mlsysauthorlist}\PaperAuthorList\end{mlsysauthorlist}
\mlsysaffiliation{epfl}{\PaperEPFL}
\mlsysaffiliation{independent}{\PaperIndependent}
\mlsyskeywords{\PaperKeywords}
\vskip 0.3in\begin{abstract}
\input{section/main/abstract}

\end{abstract}]
\printAffiliationsAndNotice{\PaperAuthorNotes}
\input{paper/main}

\label{main-content-end}\clearpage
\bibliographystyle{mlsys2025}\bibliography{bib/references}
\end{document}

%% file: paper/packages.tex
\usepackage[utf8]{inputenc}
\usepackage{amsmath,amssymb,booktabs,array,xcolor,xspace,microtype,graphicx,placeins,fix-cm,float}
\usepackage{twemojis}  %
\usepackage[most]{tcolorbox}  %
\usepackage{hyperref}
\graphicspath{{figure/generated/}{figure/source/}}
\input{cmds}

%% file: cmds.tex
\definecolor{Melon}{cmyk}{0,0.46,0.50,0}
\definecolor{BrickRed}{cmyk}{0,0.89,0.94,0.28}   %

%% file: paper/macros.tex
  \newcommand{\sysname}{CacheWeave\xspace}
  \newcommand{\sysserve}{CacheWeave-Serve\xspace}
  \newcommand{\sysshort}{CW\xspace}

\newcommand{\Kfull}{\mathcal{K}^{\mathrm{full}}} %
\newcommand{\Kpic}{\mathcal{K}^{\mathrm{PIC}}}  %
\definecolor{navy}{HTML}{1F3A68}
\newtcolorbox{takeaway}{enhanced, breakable, colback=navy!6, colframe=navy,
  boxrule=0.6pt, arc=2pt, left=5pt, right=5pt, top=3pt, bottom=3pt,
  fontupper=\small\itshape, before upper={\textbf{\upshape Takeaway.}\ }}
\makeatletter
\AtBeginDocument{\renewcommand\floatc@ruled[2]{\normalcolor{\@fs@cfont #1} #2\par}}
\makeatother

%% file: paper/metadata.tex
\newcommand{\PaperTitle}{CacheWeave: Learning to Recompute for%
\texorpdfstring{\\}{ }Efficient Position-Independent KV Caching}
\newcommand{\PaperPDFTitle}{CacheWeave: Learning to Recompute for Efficient Position-Independent KV Caching}
\newcommand{\PaperRunningTitle}{CacheWeave: Learning to Recompute for Position-Independent KV Caching}
\newcommand{\PaperAuthorName}{Yueyang Pan, Qianchao Zhu, Yuliang Liu, Teng Ma, Sanidhya Kashyap}
\newcommand{\PaperEPFL}{\'{E}cole Polytechnique F\'{e}d\'{e}rale de Lausanne}
\newcommand{\PaperIndependent}{Independent Researcher}
\newcommand{\PaperAuthorList}{%
  \mlsysauthor{Yueyang Pan}{epfl,intern}
  \mlsysauthor{Qianchao Zhu}{independent,corresponding}
  \mlsysauthor{Yuliang Liu}{independent,corresponding}
  \mlsysauthor{Teng Ma}{independent}
  \mlsysauthor{Sanidhya Kashyap}{epfl}
}
\newcommand{\PaperAuthorNotes}{%
  \textsuperscript{*}Work done during internship at Xiaohongshu Inc.\space
  \textsuperscript{\ensuremath{\dagger}}Corresponding authors.\space
}
\newcommand{\PaperCorrespondingAuthor}{Qianchao Zhu and Yuliang Liu}
\newcommand{\PaperEmail}{}
\newcommand{\PaperKeywords}{Position-independent caching, selective recomputation, KV cache reuse, learned selection, LLM serving}

%% file: section/main/abstract.tex
Position-independent caching reuses chunk KV caches across positions and
requests. Independent encoding, however, omits cross-chunk interactions and can
degrade accuracy. In our study, neither PIC training nor rule-based
recomputation, alone or combined, consistently recovers full-prefill accuracy.
Oracle-guided recomputation on PIC-trained models improves accuracy at the same
budget, motivating better selection. We present \sysname, which adapts the full
backbone to PIC, then freezes it and trains a model-native block selector on the
adapted model's oracles. The selector uses compact cached chunk features and
question hidden states without extracting backbone attention scores. It selects
contiguous token blocks aligned with KV blocks to support efficient
recomputation and limit copy-on-write amplification. Its runtime, \sysserve,
supports private and shared KV caches, keeps selection and recomputation on the
GPU, and stages sliding-window KV for hybrid attention models. Across three
models and four tasks, \sysname outperforms CacheBlend and ProphetKV by 7.2 and
3.5 percentage points on average at a 5\% recomputation budget, with a mean
relative accuracy loss of 1.6\% to full prefill. With 32-token blocks,
\sysserve cuts single-request TTFT by up to $16.1\times$ over full prefill and
$1.7\times$ over ProphetKV, and raises SLO goodput under load by up to
$7.8\times$.

%% file: paper/main.tex
\input{table/prior}

\input{section/main/introduction}
\input{section/main/background}

\input{section/main/motivation}
\input{section/main/system_design}

\input{section/main/evaluation}
\input{section/main/discussion}

\input{section/main/conclusion}

%% file: table/prior.tex
\begin{table*}[!t]
\caption{Comparison of PIC approaches. Restoration gives selected tokens access to preceding chunks; model-native selection is learned for the adapted model. Online cost is extra work beyond PIC: $\beta$ denotes the recomputed token fraction, \emph{probe} an early-layer KV deviation pass, \emph{Q pass} question prefill with attention probabilities, and \emph{selector} learned block scoring. Accuracy is relative to full prefill in our evaluation; Attuner, SemPIC, and block distillation are represented by our reader-only, writer-only, and full-model training variants.}
\label{tab:prior}
\vskip 0.15in
\centering\small
\setlength{\tabcolsep}{5pt}
\newcommand{\yes}{\twemoji{2705}}
\newcommand{\no}{\twemoji{274c}}
\begin{tabular}{@{}llcccccl@{}}
\toprule
Family & Method & Training & \shortstack{Attention\\restore} & \shortstack{Query-aware\\selection} & \shortstack{Model-native\\selection} & Online cost & Accuracy \\
\midrule
Training-free & CacheBlend & \no & \yes & \no & \no & probe $+\,\beta$ & Low \\
selective recompute & EPIC & \no & \yes & \no & \no & $\beta$ & Low \\
 & ProphetKV & \no & \yes & \yes & \no & Q pass $+\,\beta$ & Medium \\
\midrule
Training-aware & Attuner & reader & \no & -- & -- & none & Low \\
PIC (partial) & SemPIC & writer & \no & -- & -- & none & Low \\
\midrule
Training-aware PIC (full) & Block distillation & \yes & \no & -- & -- & none & Medium \\
\midrule
Native recompute & \textbf{\sysname} & \yes & \yes & \yes & \yes & selector $+\,\beta$ & High \\
\bottomrule
\end{tabular}
\end{table*}

%% file: section/main/introduction.tex
\section{Introduction}

LLM requests often reuse the same context in different combinations.
Retrieval-augmented generation (RAG) draws documents from a shared corpus, while
coding agents revisit source files and tool outputs as a task
evolves~\citep{yang2024sweagent,wang2025openhands,fang2026recache,lewis2020rag,wu2024autogen,geng2026relaycaching}.
Reusing the key-value (KV) states of these recurring chunks can avoid repeated
prefill computation. Prefix caching exploits this opportunity when requests
share an identical token prefix~\citep{zheng2024sglang,kwon2023vllm}. However, a
chunk's KV states generally depend on its preceding context. Prefix caching
therefore recomputes a recurring chunk if its preceding context changes, even
when the chunk's tokens are identical.

Position-independent caching (PIC) broadens reuse by encoding chunks
independently and assembling their cached KV for each
request~\citep{gim2024promptcache,lu2024turborag,hu2025epic}. This avoids
repeated chunk prefill but removes cross-chunk interactions from the cached
representations, which can hurt tasks that combine evidence across chunks.
Existing remedies follow two approaches (\autoref{tab:prior}). Training-free
methods selectively recompute tokens using predefined criteria, such as KV
deviation or question attention, without changing model
weights~\citep{yao2025cacheblend,wang2026prophetkv}. Training-aware methods
instead adapt how the model constructs or consumes independent chunk caches,
avoiding per-request chunk
recomputation~\citep{ma2025blockattention,chen2026attuner,xie2026sempic}.

Our study finds that neither PIC training nor selective recomputation
consistently recovers the base model's full-prefill accuracy across the models
and tasks we evaluate (\autoref{sec:motivation}). Applying ProphetKV or
CacheBlend to PIC-trained models also yields inconsistent gains and can reduce
multi-hop QA accuracy. Yet at the same 5\% recomputation budget, oracle
selectors on these models improve over ProphetKV by up to 9.7 percentage points.
These oracles require information unavailable during normal PIC inference, but
demonstrate room for better selection at a fixed budget. This motivates learning
a selector from the adapted model's own oracles.

Better selection must also preserve PIC's serving advantage. Extracting the
backbone's attention probabilities can require materializing matrices that fused
attention kernels normally avoid storing~\citep{dao2022flashattention}. In our
Qwen3-4B measurements at 16K context, ProphetKV's selection and 5\%
recomputation reduce warm PIC's speedup over full prefill from $14\times$ to
$3.1\times$ (\autoref{sec:motivation-efficiency}). Updates also affect memory
sharing: in a block-based KV cache~\citep{kwon2023vllm}, recomputing scattered
tokens can require copying many shared blocks. A small token budget can
therefore incur substantial execution and memory costs. Effective recomputation
requires choosing both which states to update and how to execute and store those
updates.

We present \sysname, which couples PIC adaptation with a learned, model-native
recomputation selector. Using cached chunk features and question hidden states,
the selector chooses contiguous token blocks aligned with the engine's KV
blocks, without extracting backbone attention probabilities
(\autoref{sec:selector}). We first fine-tune the full backbone under PIC by
self-distillation on the base model's full-prefill responses, then freeze it and
train the selector on oracles computed from that adapted backbone
(\autoref{sec:training}). For a fixed block size, one selector supports multiple
recomputation budgets without retraining. Our runtime, \sysserve, uses block
alignment to bound copy-on-write amplification, supports private and shared GPU
caches, and keeps selection and recomputation on the GPU. It also supports
hybrid attention models by staging sliding-window KV layer by layer
(\autoref{sec:serving}).

We evaluate \sysname on three models from two families and four tasks covering
multi-hop QA and retrieval (\autoref{sec:evaluation}). At a 5\% recomputation
budget, \sysname outperforms CacheBlend and ProphetKV on the same PIC-trained
models over all model-task pairs, gaining 7.2 and 3.5 points on average. Its
relative accuracy loss to full prefill averages 1.6\% and is at most 8.3\%; its
average lead over ProphetKV persists at every tested block size. At 24K material
tokens, with 32-token blocks as recompute granularity and a 5\% budget,
\sysserve achieves single-request time-to-first-token (TTFT) speedups of
$13.9\times$ over full prefill and $1.46\times$ over ProphetKV on Qwen3-32B, and
$16.1\times$ over full prefill on Gemma4-12B. Under load, \sysserve raises peak
SLO goodput by up to $7.8\times$.

\noindent We make the following contributions:
\par

\begin{samepage}
\begin{itemize}
    \item A diagnosis of PIC accuracy and cost: either PIC training or
    rule-based recomputation stays far from full-prefill accuracy. Simple
    combination can still leave the gap wide. Oracle selections at the same
    budget do much better, and inefficient recomputation can consume a lot of
    PIC's prefill speedup (\autoref{sec:motivation}).

    \item A model-native selector and a two-stage recipe that trains it on the
    PIC-trained model's own oracles, with no attention scores at inference and
    selector granularity aligned with KV block sizes
    (\autoref{sec:selector}--\ref{sec:training}).

    \item \sysserve, a serving design providing private and shared GPU memory
    modes with an analysis of when sharing pays, efficient GPU-resident
    selection and recomputation, and support for hybrid sliding-window models
    (\autoref{sec:serving}).

    \item An evaluation on three models from two families and four tasks,
    demonstrating higher accuracy at matched recomputation budgets and
    substantial single-request TTFT reductions at long contexts
    (\autoref{sec:evaluation}).
\end{itemize}
\end{samepage}

%% file: section/main/background.tex
\section{Background and Related Work}
\label{sec:background}

\begin{figure}[!t]
\centering
\includegraphics[width=\columnwidth]{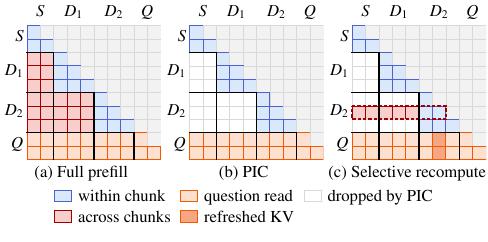}
\caption{Attention patterns for system prefix $S$, chunks $D_1,D_2$,
and question $Q$; rows attend to coloured columns.
(a) Full prefill. (b) PIC removes cross-chunk interactions.
(c) Selective recomputation restores selected tokens' access to
preceding context (dashed), and $Q$ reads their refreshed KV.
Training-aware PIC retains (b) and adapts the model's weights.}
\label{fig:pic-keysets}
\end{figure}

\paragraph{Reusable context across requests.}
RAG systems and tool-using agents assemble requests from reusable material, such
as documents, tool descriptions, and tool
outputs~\citep{yang2024sweagent,wang2025openhands,lewis2020rag,wu2024autogen,izacard2020fid}.
We call each unit a \emph{chunk}. The same chunk may recur across requests in a
different order or after a different history. We write an assembled request as
\begin{equation}
  X = (S, D_1, D_2, \dots, D_n, Q),
  \label{eq:request}
\end{equation}
where $S$ is a shared system prefix, $D_i$ are material chunks, and $Q$ is the
request-specific question, followed by the answer $A$. Let $\sigma(t)
\in \{S, D_1, \ldots, D_n, Q\}$ denote the span containing input token $t$, and
$p_i$ the starting position of $D_i$.
\paragraph{Beyond prefix caching.}
Under \emph{full causal prefill}, token $t$ attends to 
\begin{equation}
  \Kfull(t) = \{u : u \le t\},
  \label{eq:full-keys}
\end{equation}
so a chunk's KV states generally depend on its preceding context.
Prefix caching therefore reuses states only when the entire token prefix
matches~\citep{zheng2024sglang,kwon2023vllm}. \emph{Position-independent
caching} broadens reuse by encoding each chunk
independently~\citep{gim2024promptcache,lu2024turborag,yao2025cacheblend}:
\begin{equation}
  \Kpic(t) =
  \begin{cases}
    \{u \le t : \sigma(u) = \sigma(t)\} & \sigma(t) \ne Q,\\
    \{u : u \le t\} & \sigma(t) = Q.
  \end{cases}
  \label{eq:pic-keys}
\end{equation}
At serving time, cached chunks are assembled at their request
positions, and only $Q$ is prefilled (\autoref{fig:pic-keysets}a--b).
With rotary position embeddings, a cached key $k_j$ at chunk-local
position $j$ is relocated as
$\widetilde{k}_{p_i+j}=R_{p_i}k_j$, leaving its value unchanged.
Position alignment leaves cross-chunk interactions missing, causing substantial
accuracy loss on tasks requiring reasoning across chunks. Prior work mitigates
this loss in two ways (\autoref{tab:prior}): \emph{training-free selective
recomputation} restores some of the missing interactions at inference time,
while \emph{training-aware PIC} trains the model to use independently encoded
caches.

\begin{figure*}[!t]
\centering
\includegraphics[width=\textwidth]{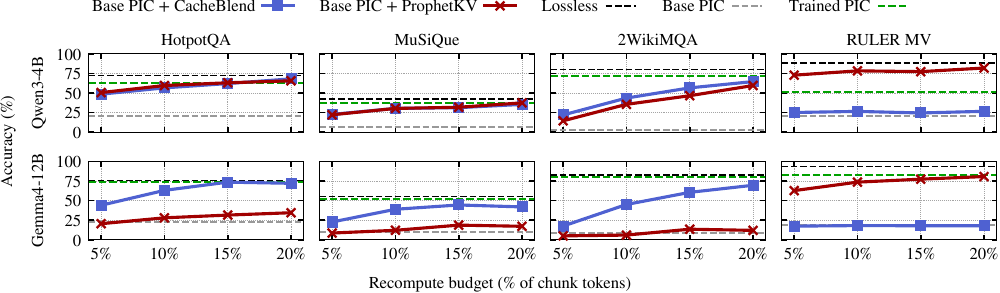}
\caption{Each PIC method alone. Lines: non-native recomputation on the untrained
base model at 5--20\% recomputation budget. Dashed: lossless (base model, full
prefill), base PIC, and trained PIC without recomputation (green). Trained PIC
still trails lossless by 36.9 (Qwen3-4B) and 10.7 (Gemma4-12B) points on MV,
and by at most 13.8\% of lossless on multi-hop QA.}
\label{fig:motivation-base}
\end{figure*}

\paragraph{Training-free selective recomputation.}
These methods recompute a small set $\mathcal{R}$ of chunk tokens
with access to their full preceding context, while retaining PIC
for the rest (\autoref{fig:pic-keysets}c):
\begin{equation}
  \mathcal{K}^{\mathcal{R}}(t) =
  \begin{cases}
    \Kfull(t) & t \in \mathcal{R},\\
    \Kpic(t) & \text{otherwise,}
  \end{cases}
  \label{eq:recompute-keys}
\end{equation}
where $|\mathcal{R}| \le \beta \sum_i |D_i|$ limits the recomputed
fraction. Selected tokens are recomputed against the reused cache,
and the question reads their refreshed KV.

The methods differ in how they select $\mathcal{R}$.
\emph{Query-agnostic} methods choose it without seeing the query:
CacheBlend~\citep{yao2025cacheblend} and KVShare~\citep{yang2025kvshare}
recompute the tokens whose KV deviates most between PIC and full prefill in
early layers; EPIC~\citep{hu2025epic} recomputes the first tokens of each chunk;
RedKnot~\citep{liu2026redknot} recomputes selected attention heads instead of
tokens; and Cache-Craft~\citep{agarwal2025cachecraft} couples targeted
recomputation with chunk-cache management. \emph{Query-aware} methods, such as
ProphetKV~\citep{wang2026prophetkv}, InfoFlow KV~\citep{teng2026infoflowkv},
QCFuse~\citep{yan2026qcfuse} and RelaxKV~\citep{qi2026relaxkv}, choose
$\mathcal{R}$ from the query's attention to the chunks. All of them leave the
model unchanged, choose $\mathcal{R}$ by a fixed rule, and trade accuracy
against the recomputation budget $\beta$.

\paragraph{Training-aware PIC.}
These methods train the model to use independently encoded chunks
without per-request KV recomputation.
Block-Attention~\citep{ma2025blockattention} trains the whole model by
supervised fine-tuning (SFT) under PIC prefill, and block
distillation~\citep{li2026blockdistillation} further distills from a
full-attention teacher. Others add trainable components:
KVLink~\citep{yang2025kvlink} trains link tokens jointly with the model, KV
Packet~\citep{chen2026kvpacket} wraps each cached chunk in trainable boundary
embeddings, and COMB~\citep{zhao2026comb} adds a trained chunk encoder and
cross-attention interface to a frozen decoder. Two recent methods train one side
of the interaction: SemPIC~\citep{xie2026sempic} adapts the writer side, how
chunks are encoded, and Attuner~\citep{chen2026attuner} adapts the reader side,
how the question reads fixed caches. 
These approaches improve cache construction or consumption but
leave cached chunk KV unchanged in the request-specific context.

\paragraph{Serving costs of PIC.}
Accuracy recovery must preserve PIC's latency and memory benefits.
Selection adds runtime overhead, especially when it requires attention scores
that fused kernels~\citep{dao2022flashattention} do not expose. Meanwhile, KV
allocation and sharing determine batch capacity~\citep{kwon2023vllm};
request-specific recomputation can reduce the memory savings from sharing
through copy-on-write. Efficient PIC therefore requires coordinating cache
management, data movement, and selective recomputation.

%% file: section/main/motivation.tex
\section{Motivation for Learned Recomputation}
\label{sec:motivation}
We examine the accuracy and cost of PIC training and selective
recomputation, separately and in combination, using the models and
benchmarks in \autoref{sec:eval-setup}.
We use the base model's full-prefill accuracy as the \emph{lossless}
reference and denote PIC-trained inference without recomputation
as \emph{trained PIC}.
For selective recomputation, we evaluate
ProphetKV~\citep{wang2026prophetkv} and
CacheBlend~\citep{yao2025cacheblend}.
We refer to their selection policies as \emph{non-native}:
they use predefined rules rather than selectors trained for
the PIC-adapted model.

\subsection{Each Family Alone Leaves Accuracy Gaps}
\label{sec:motivation-gap}
We first measure PIC training without recomputation and selective
recomputation on the base model (\autoref{fig:motivation-base}).
PIC training alone leaves accuracy gaps that vary across models and tasks.
On the RULER MV retrieval task, trained PIC still loses 36.9 points on Qwen3-4B
and 10.7 on Gemma4-12B. On multi-hop QA, the gap is small for Gemma4-12B (2.6\%
to 6.1\% of lossless) but remains larger for Qwen3-4B, up to 13.8\% (36.7
against 42.6 on MuSiQue). Even after PIC training, the model thus cannot make up
for the missing cross-chunk attention in every setting.

Selective recomputation also leaves substantial gaps, even with a large
budget. On 2WikiMQA, base Qwen3-4B scores 2.5 under PIC against 80.0 lossless;
recomputing 20\% of the chunk tokens raises it to 59.5 with ProphetKV and 64.5
with CacheBlend, still 15 to 21 points short. How much each method helps also
depends on the model and the task. On Gemma4-12B, ProphetKV helps multi-hop QA
little: at 20\% it gains only 3 to 12 points over base PIC. CacheBlend, in turn,
stays far from lossless on MV for both models, reaching 26.6 and 17.9 at
20\% against 88.6 and 93.5 lossless.

\begin{takeaway}
Neither PIC training nor selective recomputation alone consistently
recovers full-prefill accuracy, and their remaining gaps vary across
models and tasks. This motivates us to examine whether combining
them closes these gaps.
\end{takeaway}

\subsection{Combining Them Gives Inconsistent Gains}
\label{sec:motivation-compose}

\begin{figure}[!t]
\centering
\includegraphics[width=\columnwidth]{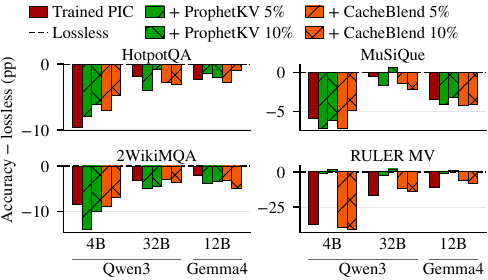}
\caption{Non-native recomputation on PIC-trained models: accuracy minus
lossless (percentage points; three draws). Trained PIC reads the PIC cache
without recomputation; ProphetKV and CacheBlend recompute 5\% or 10\% of the
chunk tokens.}
\label{fig:motivation-gap}
\end{figure}

We next apply ProphetKV and CacheBlend to the three PIC-trained
models to test whether recomputation closes the gaps left by
training (\autoref{fig:motivation-gap}).

ProphetKV improves retrieval but provides inconsistent gains on multi-hop QA.
Pooled over the three models, at 5\% it gains 19.9 points over trained PIC on
MV. On 2WikiMQA it makes answers worse: at 5\% it is 2.9 points below trained
PIC (95\% CI $[-3.3,-0.8]$), and all three models score below their trained-PIC baselines. Even on Qwen3-4B, which
keeps the largest multi-hop loss after training, ProphetKV at 5\% changes
HotpotQA, MuSiQue and 2WikiMQA by $+1.7$, $-1.2$ and $-5.5$ points respectively.

CacheBlend offers limited gains even on MV retrieval: even at 10\%, it changes
accuracy by only $-4.0$ to $+3.2$ points across models, leaving the gap to
lossless wide. It never improves multi-hop QA significantly at any budget;
its pooled change stays between $-1.0$ and $+1.7$ points. Stacking a non-native
method on a PIC-trained model thus does not reliably recover the loss
training leaves on multi-hop QA.

\begin{takeaway}
Non-native recomputation does not reliably improve PIC-trained models
and can even reduce accuracy on tasks requiring cross-chunk reasoning.
These results motivate a closer examination of token selection.
\end{takeaway}

\subsection{Better Selection Improves Recovery}
\label{sec:motivation-oracle}
\begin{figure}[!t]
\centering
\includegraphics[width=\columnwidth]{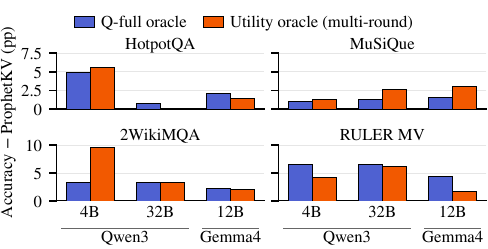}
\caption{Task accuracy of two oracle selectors minus ProphetKV, all at a 5\%
budget on the same PIC-trained model (percentage points; three draws).}
\label{fig:motivation-selection}
\end{figure}

Do these limitations come from recomputation itself or from which tokens are
selected? We compare ProphetKV with two oracle selectors at the same 5\% budget.
Both oracles are infeasible in deployment; they bound what a better selector
could achieve.

\paragraph{Full-prefill attention oracle (Q-full).} 
The \emph{Q-full} oracle applies ProphetKV's selection rule to full-prefill KV
rather than PIC states.
Let $\mathcal{C}$ be the chunk tokens, $\mathcal{L}$ the full-attention
layers (our ProphetKV extension excludes sliding-window layers), and
$a^{\ell,h}_{r,t}$ the attention of question token $r\in Q$ to chunk token $t$
in head $h$ of layer $\ell$ under full prefill. The score of $t$ averages
its attention, renormalized over the chunk tokens, over question tokens, heads
and layers:
\begin{equation}
  s(t) = \frac{1}{|\mathcal{L}|}\sum_{\ell\in\mathcal{L}}\frac{1}{H}\sum_{h=1}^{H}
  \frac{1}{|Q|}\sum_{r\in Q}
  \frac{a^{\ell,h}_{r,t}}{\sum_{t'\in\mathcal{C}} a^{\ell,h}_{r,t'}},
  \label{eq:qfull}
\end{equation}
and the top $\lceil\beta|\mathcal{C}|\rceil$ tokens are recomputed. ProphetKV
uses the same score method against the PIC KV cache. Q-full is
infeasible in deployment because it requires the full prefill.

\paragraph{Multi-round utility oracle.} 
The \emph{utility oracle} estimates how replacing a token's cached KV with its
full-prefill KV would affect the likelihood of the full-prefill answer $y$,
generated by the same PIC-trained model:
\begin{equation}
\begin{split}
u(t)=\sum_{\ell}\Bigl(&\bigl\langle \nabla_{K_\ell(t)}\log p(y),\,\Delta K_\ell(t)\bigr\rangle\\
&+\bigl\langle \nabla_{V_\ell(t)}\log p(y),\,\Delta V_\ell(t)\bigr\rangle\Bigr),
\end{split}
\label{eq:utility}
\end{equation}
where $\Delta K_\ell(t)=K^{\mathrm{full}}_\ell(t)-K_\ell(t)$, $\Delta
V_\ell(t)$ likewise, and $\langle \cdot, \cdot \rangle$ is the inner
product. The gradient measures how sensitive the answer is to the token's KV,
and the difference measures how wrong the reused KV is; their product estimates
how much recomputing the token would help. Because tokens interact, the estimate
is iterative: selection is greedy over rounds of equal size, and each round
recomputes the tokens chosen so far exactly and re-linearizes
\autoref{eq:utility} at the new state. We empirically use eight rounds for 10\%
budget, and the first four rounds form the set for 5\%. The utility oracle is
also infeasible in deployment because it requires both the full-prefill answer
and KV cache.

\paragraph{Results.} \autoref{fig:motivation-selection} compares the oracles
with ProphetKV; we omit CacheBlend, which trails ProphetKV
(\autoref{sec:motivation-compose}). Both oracles match or beat ProphetKV on
every model and task, by up to 9.7 points (utility on Qwen3-4B, 2WikiMQA). Since
Q-full, like ProphetKV, sees only the question, part of ProphetKV's loss comes
from scoring tokens on the PIC cache rather than the full-prefill state.

\begin{takeaway}
Better selection improves recovery without increasing the
recomputation budget. These oracles provide supervision for learning a selector
tailored to the PIC-trained model.
\end{takeaway}

\subsection{Selection and Recomputation Must Be Cheap}
\label{sec:motivation-efficiency}
\begin{figure}[!t]
\centering
\includegraphics[width=\columnwidth]{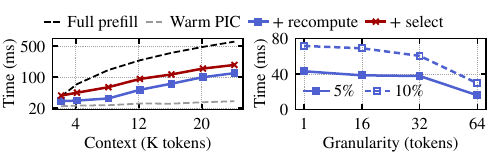}
\caption{ProphetKV recomputation overhead on Qwen3-4B with 1024-token chunks.
Left: warm PIC (question-only prefill), plus 5\% recomputation,
then selection. Right: recomputation latency versus granularity
at 16K context, with 5\% and 10\% budgets.}
\label{fig:motivation-overhead}
\end{figure}

Accuracy recovery comes at a runtime cost.
\autoref{fig:motivation-overhead} measures the cost of ProphetKV on Qwen3-4B. At
16K context, on top of warm PIC's 25\,ms, selection alone adds 43\,ms, because
ProphetKV needs the question's attention score, which fused attention
kernels do not expose. Recomputing 5\% of the chunk tokens adds another 46\,ms.
Together, these overheads reduce the speedup over
full prefill from $14\times$ to $3.1\times$.
Recomputation cost depends on budget and granularity.
Increasing the budget from 5\% to 10\% raises recomputation latency by
$1.7\times$, while grouping selected tokens into contiguous runs reduces latency
at either budget.

\begin{takeaway}
Selection and recomputation can consume most of PIC's prefill speedup. 
Preserving this speedup requires low-overhead selection
and a small recomputation budget spent on efficiently processed
token blocks.
\end{takeaway}

%% file: section/main/system_design.tex
\section{\sysname Design}
\label{sec:design}

\sysname combines PIC adaptation with a model-native block selector
(\autoref{sec:selector}). We first adapt the backbone, then train the
selector on the adapted model's oracles (\autoref{sec:training}).
At inference, selection uses cached chunk features and question hidden
states without extracting attention probabilities. \sysserve coordinates
block-aligned recomputation and cache sharing to limit copying and
preserve PIC's speedup (\autoref{sec:serving}).

\begin{figure}[!t]
\centering
\includegraphics[width=\columnwidth]{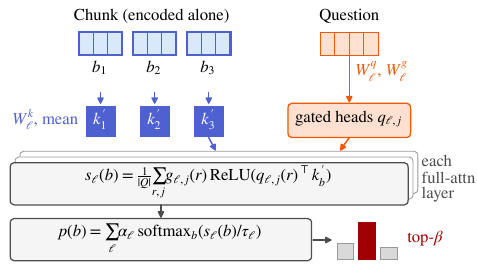}
\caption{Block scoring at one layer: offline writer keys are scored by
question-derived reader heads. Learned weights combine layer distributions to
select blocks for recomputation.}
\label{fig:selector}
\end{figure}

\subsection{Model-Native Selector}
\label{sec:selector}
\paragraph{Selector granularity.} 
We partition chunks into contiguous \emph{selector blocks} of $g$ tokens.
Selector and KV blocks share a chunk-local origin, with $g$ dividing the KV
block size $B$ ($B=64$), so each selector block touches at most one KV block.
From selector blocks $\mathcal{G}$, we select $R\subseteq\mathcal{G}$ with
$|R|=\lceil\beta|\mathcal{G}|\rceil$ for budget $\beta$ and recompute their
tokens with preceding context (\autoref{eq:recompute-keys}). The question then
reads the refreshed KV. Alignment limits both scattered computation and
copy-on-write amplification.

\paragraph{Offline block key construction.}
A writer projection maps independently computed chunk hidden states
$h^{\mathrm{pic}}_\ell(t)$ to $d_{\mathrm{idx}}=128$ dimensions and averages
them within each block $b$:
\begin{equation}
  k'_\ell(b) = \frac{1}{|b|}\sum_{t\in b} W^{k}_\ell\,h^{\mathrm{pic}}_\ell(t).
  \label{eq:block-key}
\end{equation}
These keys are stored with the chunk's KV cache (\autoref{fig:selector}).

\paragraph{Online learned block scoring.}
Question prefill over PIC KV produces hidden states $h_\ell(r)$, from which
the reader derives $H=4$ heads $q_{\ell,j}(r)=W^q_{\ell,j}h_\ell(r)$ and
gates $g_\ell(r)=\operatorname{softmax}_j(W^g_\ell h_\ell(r))$:
\begin{align}
  s_\ell(b) &= \frac{1}{|Q|}\sum_{r\in Q}\sum_{j=1}^{H}g_{\ell,j}(r)
  \operatorname{ReLU}\bigl(q_{\ell,j}(r)^\top k'_\ell(b)\bigr),
  \label{eq:block-score}\\
  p(b) &= \sum_{\ell\in\mathcal{L}}\alpha_\ell
  \operatorname{softmax}_{b}\bigl(s_\ell(b)/\tau_\ell\bigr),
  \label{eq:layer-mix}
\end{align}
where $\alpha=\operatorname{softmax}(\gamma)$ and the temperatures $\tau_\ell$
are learned. $\mathcal{L}$ contains all full-attention layers: 36 on Qwen3-4B,
64 on Qwen3-32B, and 8 on Gemma4-12B.

\paragraph{Selector overhead.}
Selection uses hidden states and cached keys, requiring no attention
scores or changes to fused attention kernels. Compared with
ProphetKV, query--key scoring uses $256\times$ and $512\times$ fewer scalar
multiplications on Qwen3-4B and Qwen3-32B at $g=32$, excluding reader
projections and other costs. Both methods use 128-dimensional scoring heads
on these models, but the selector scores blocks with four heads rather than
tokens with 32 or 64 heads.
For full blocks on Qwen3-4B, selector keys add 144 values per token at
$g=32$, about 0.2\% of KV storage at equal precision.
Parameter overhead is 1.5\% on Qwen3-4B, 0.6\% on Qwen3-32B, and 0.2\%
on Gemma4-12B.

\subsection{Two-Stage Training}
\label{sec:training}

\paragraph{Stage 1: PIC self-distillation.}
We generate target responses with the base model under full prefill, then
fine-tune the entire backbone under the PIC mask (\autoref{eq:pic-keys})
using cross-entropy loss. The model thus learns to predict
full-context targets from independently encoded chunk caches, without
external reference answers. Updating all backbone parameters jointly
adapts cache encoding and reading. Our ablation supports this choice:
on Qwen3-4B's 2WikiMQA, full-backbone PIC-SFT scores 71.8, compared with
13.5 and 61.5 for our reader-only
(Attuner-like~\citep{chen2026attuner}) and writer-only
(SemPIC-like~\citep{xie2026sempic}) variants.

\paragraph{Stage 2: oracle-supervised selection.}
We freeze the adapted backbone and train the selector using oracles from that
backbone (\autoref{sec:motivation-oracle}): utility for multi-hop QA and Q-full
for retrieval, chosen on the development set. For oracle block distribution
$p^*$ and prediction $p$, the loss is
\begin{equation}
  \mathcal{J}=\mathrm{KL}(p^*\,\Vert\,p)
  +\mathrm{KL}(p^*_U\,\Vert\,p_U),
  \label{eq:selector-loss}
\end{equation}
where the subscript $U$ denotes renormalization over the union of the
oracle's and selector's top $2m$ blocks.
For each sample, we draw $\beta\sim\mathcal{U}(0.025,0.15)$ and set
$m=\lceil\beta|\mathcal{G}|\rceil$. The first term matches the full
distribution; the second emphasizes ranking near the selection cutoff.
Varying the budget allows one selector to serve different budgets without
retraining. Its chunk and question inputs come from the frozen backbone's
PIC hidden states, so gradients update only the selector.

\subsection{Efficient PIC Serving}
\label{sec:serving}
\begin{figure}[!t]
\centering
\includegraphics[width=\columnwidth]{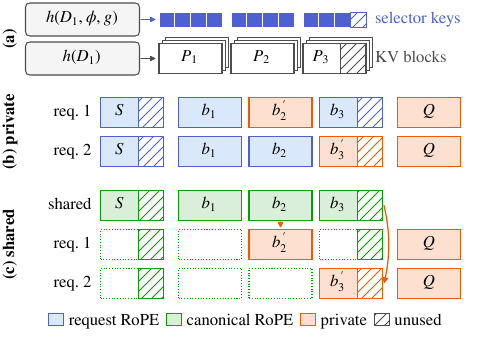}
\caption{Cache layout and HBM modes. (a) Chunk IDs index KV blocks and selector
keys. (b) Private copies updated in place. (c) Shared blocks copied before
updates; unchanged blocks read from the common base (dotted).}
\label{fig:serving-view}
\end{figure}

\sysserve executes the adapted backbone and selector over a two-tier cache.
Host DRAM or a memory pool (L2) stores immutable chunk KV and selector keys
(\autoref{fig:serving-view}a), while GPU HBM (L1) holds resident objects for
active requests~\citep{liu2025lmcache,qin2024mooncake}. For each request,
\sysserve reuses resident objects or loads missing ones, selects and recomputes
blocks, then decodes from the refreshed cache.

\subsubsection{Cached Objects and Request Layout}
\label{sec:serving-layout}
Requests supply tokens and content hashes for chunks $D_1,\dots,D_n$.
For a fixed backbone and cache format, \sysserve indexes chunk KV and selector
keys as
\begin{align}
  h(D) &\mapsto [P_1,\dots,P_{\lceil |D|/B\rceil}],
  \label{eq:kv-entry}\\
  h(D,\phi,g) &\mapsto
  \{k'_{\ell,1},\dots,k'_{\ell,\lceil |D|/g\rceil}\}_{\ell\in\mathcal{L}},
  \label{eq:selector-entry}
\end{align}
where $h$ hashes its arguments and $\phi$ identifies the selector. 
Each $P_j$ holds KV for up to $B$ consecutive tokens, with attention keys stored
before RoPE for reuse at different positions. The final block has at most $B-1$
unused slots, which the request's token-to-slot map excludes from attention.
Matched KV is registered in the prefix tree for reuse on subsequent turns.

\subsubsection{Private and Shared KV in HBM}
\label{sec:serving-modes}
In \emph{private} mode, each request owns chunk copies and recomputes selected
tokens in place. In \emph{shared} mode, requests read common blocks and copy any
block still read by another request before updating it. Copy-on-write operates
on entire KV blocks, even when only a few tokens within a block are recomputed.

\paragraph{Granularity and memory savings.}
Consider $r$ concurrent requests, each with $N$ cached tokens eligible for
sharing; a fraction $s$ is common to all requests and the rest unique. We count
resident material KV in token slots, excluding block rounding, reserved buffers,
and non-material state. Let $cN$ be the mean shared KV copied per request.
Selecting $\beta N/g$ aligned blocks gives $c\le\min(s,\beta B/g)$. The two
modes require
\begin{align}
  M_{\mathrm{private}} &= rN,
  \label{eq:mem-private}\\
  M_{\mathrm{shared}} &= \bigl(s+r(1-s)+rc\bigr)N,
  \label{eq:mem-shared}
\end{align}
where shared mode stores one common base, request-specific material, and private
replacements. Sharing saves material KV when $c<s(1-1/r)$. For token-level
selection ($g=1$), $B=64$, and $\beta=5\%$, the bound $c\le s$ cannot guarantee
savings. ProphetKV touches 61--91\% of all KV blocks on our models; actual
copying depends on which are shared. With $g=B/2$, copied capacity is at most
twice the selected token count, giving $c\le2\beta$. For $s=0.6$ and
$\beta=5\%$, the worst-case $c=0.1$ yields
$M_{\mathrm{private}}/M_{\mathrm{shared}}=r/(0.6+0.5r)$, approaching $2\times$
as concurrency grows; fewer copies improve this ratio.

\paragraph{Sharing overhead.}
Shared HBM keys use chunk-local RoPE positions $0,\dots,|D|-1$.
For an object at offset $o$, a query at position $p_q$ uses rotation
$R_{p_q-o}$; per-object attention outputs are combined using their log-sum-exp
normalizers. This position handling applies to every attention call, including
decode. Copy-on-write adds another cost, with worst-case copy capacity reserved
at admission before selection. Private mode instead rotates keys to request
positions when loading them and updates its copies in place. It avoids these
sharing costs but duplicates reused KV across requests. \sysserve defaults to
private mode; sharing can benefit larger batches with substantial common
material when memory savings outweigh these costs.

\subsubsection{GPU-Resident Selection and Recomputation}
\label{sec:efficient-selection-recomputation}

\paragraph{Selection.}
Both steps stay on the GPU, with no host round trip between selection and
recomputation (Algorithm~\ref{alg:serving}). Each chunk's selector keys across
full-attention layers are packed into one host buffer for a single transfer. On
Qwen3, transfer overlaps question prefill, and reader projections are batched
after prefill; on Gemma4, projections are batched within each full-attention
layer. A fused kernel applies ReLU, head gating, and the reduction in
\autoref{eq:block-score}.

\paragraph{Recomputation.}
Selected blocks form short query segments over cached prefixes.
On Qwen3, contiguous runs use variable-length FlashAttention-3 at their original
request positions. Gemma4 uses FlashAttention-3 for sliding-window attention
(SWA) over staged windows and the question's SWA KV (\autoref{sec:hybrid-swa}),
and FlashAttention-2 for full-attention layers.

\begin{algorithm}[t]
\caption{Two-step prefill in \sysserve.}
\label{alg:serving}
\begin{algorithmic}[1]
\REQUIRE objects needed for Step 1 loaded; buffers reserved at admission
\STATE \emph{Step 1:} prefill $Q$ over PIC KV; compute block scores $p$
(\autoref{eq:block-score}--\ref{eq:layer-mix})
\STATE $R \gets$ top-$\lceil\beta|\mathcal{G}|\rceil$ selector blocks by $p$
\STATE $\mathcal{P} \gets$ KV blocks touched by $R$
\IF{shared mode}
  \STATE copy blocks in $\mathcal{P}$ still read by other requests;
  remap this request to the private copies
\ENDIF
\STATE \emph{Step 2:} recompute tokens covered by $R$ and prefill $Q$
over the updated cache view, layer by layer, at their original positions
\RETURN output of the last token of $Q$
\end{algorithmic}
\end{algorithm}

\subsubsection{Cache Management for Hybrid Attention}
\label{sec:hybrid-swa}
For hybrid models such as Gemma4, \sysserve keeps cached full-attention KV
in HBM and stages SWA KV on demand.
Before selection, it loads only the SWA window preceding $Q$; question
states are computed during prefill. Interior recomputation may require
older windows, so all chunk SWA KV is retained in L2. After selection,
the required windows are staged layer by layer, overlapping transfers
with recomputation. Full-attention KV can be shared across requests,
whereas staged SWA KV remains private. This limits HBM residency
while preserving the complete SWA cache in L2 (\autoref{sec:discussion}).

%% file: section/main/evaluation.tex
\section{Evaluation}
\label{sec:evaluation}
We evaluate the accuracy of \sysname and the efficiency of \sysserve across
datasets, model architectures and request patterns. Our key findings are:
\begin{itemize}
  \item \textbf{Higher quality} (\autoref{sec:eval-accuracy}). At a 5\% budget,
  \sysname outperforms CacheBlend and ProphetKV on all 12 model--task pairs,
  by 7.2 and 3.5 points on average, with at most 8.3\% relative loss to full
  prefill. It also leads ProphetKV on all 12 pairs at $g=32$.
  \item \textbf{Lower TTFT} (\autoref{sec:eval-serving}). At 24K context,
  \sysserve cuts single-request TTFT on Qwen3-32B by 13.9$\times$ over full
  prefill and 1.46$\times$ over ProphetKV, and by 16.1$\times$ on the hybrid
  Gemma4-12B.
  \item \textbf{Higher throughput} (\autoref{sec:eval-serving}). \sysserve
  raises peak SLO goodput over the best baseline by up to 7.8$\times$ under
  low sharing and 7.6$\times$ under high sharing.
\end{itemize}

\subsection{Experimental Setup}
\label{sec:eval-setup}
\begin{figure*}[!t]
\centering
\includegraphics[width=\textwidth]{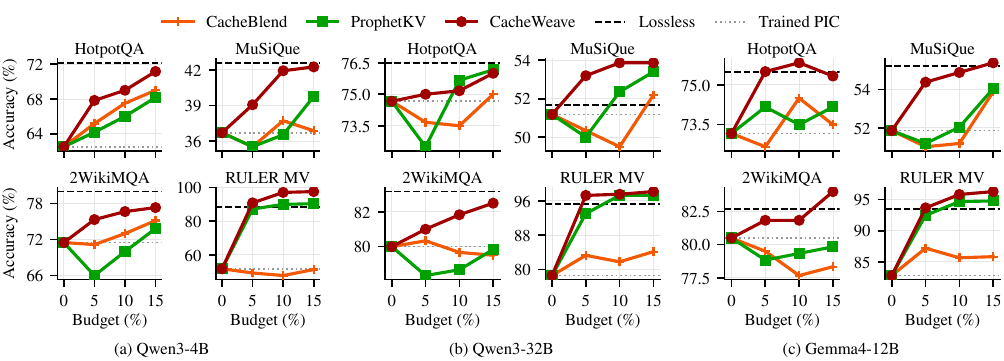}
\caption{Accuracy on the PIC-trained models against the recomputation budget,
with token-level selection ($g=1$). Every method starts from trained PIC at
0\%. At 5\%, \sysname loses at most 8.3\% relative to lossless, against 44.5\%
for CacheBlend and 17.5\% for ProphetKV, and it never falls below trained PIC.}
\label{fig:eval-accuracy}
\end{figure*}
\noindent\textbf{Models.} We evaluate three models from two families: Qwen3-4B
and 32B~\citep{qwen2025qwen3}, with full attention only, and
Gemma4-12B~\citep{gemma2026gemma4}, a hybrid model with 8 full-attention layers
and 40 SWA layers with a 1{,}024-token window. Each model is PIC-trained with
its own selector (\autoref{sec:training}).

\noindent\textbf{Workloads and metrics.} Following prior work, we use four tasks
in two categories. \emph{Multi-hop QA (DocQA):} HotpotQA, MuSiQue and
2WikiMultihopQA~\citep{yang2018hotpotqa,trivedi2022musique,ho2020twowiki} from
LongBench~\citep{bai2024longbench} require reasoning across passages. Each
passage forms one chunk, and a prompt holds 2 to 20 passages of evidence and
distractors, up to 18K tokens. \emph{Retrieval:} RULER multi-value
(MV)~\citep{hsieh2024ruler} asks for all values of a key hidden in 2 to 12
chunks, up to 24K tokens; on the other RULER tasks trained PIC is already
lossless. An LLM (DeepSeek-V4.1-Flash) judges QA accuracy, and F1 gives the
same ranking; MV uses RULER's scorer. We report the mean of three samples.

\noindent\textbf{Baselines.} For accuracy, we compare with
CacheBlend~\citep{yao2025cacheblend} and ProphetKV~\citep{wang2026prophetkv} on
the same PIC-trained models and budgets, with full prefill on the base model
(lossless) and trained PIC without recomputation as references. We omit
EPIC~\citep{hu2025epic}, whose accuracy is close to or below CacheBlend's. On
Gemma4-12B, we adapt ProphetKV to score with the full-attention layers only, due
to the window limit of SWA; for $g>1$ it ranks blocks by their mean token
score. For efficiency, we build \sysserve on SGLang~\citep{zheng2024sglang} and
compare with full prefill and prefix caching in the same engine, with ProphetKV
ported into \sysserve, and with CacheBlend's own implementation on
LMCache~\citep{liu2025lmcache}.

\noindent\textbf{Platform.} We train and evaluate on 8 NVIDIA H200 GPUs with two
48-core Intel Xeon Platinum 8558 CPUs (192 threads) and 2TB DRAM. Unless noted,
\sysserve uses KV blocks of $B=64$ tokens.

\subsection{Accuracy}
\label{sec:eval-accuracy}
\begin{figure}[!h]
\centering
\includegraphics[width=\columnwidth]{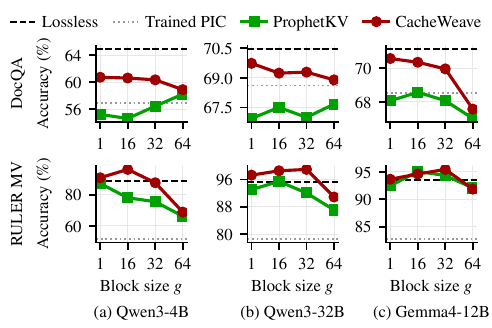}
\caption{Accuracy against selector block size at a 5\% budget. DocQA averages
HotpotQA, MuSiQue and 2WikiMultihopQA equally; CacheBlend omitted for small
gains over trained PIC.}
\label{fig:eval-granularity}
\end{figure}
\autoref{fig:eval-accuracy} compares the methods at 5--15\% budgets with
token-level selection ($g=1$), as in prior work. \sysname has the best
four-task average on every model at every budget. At 5\%, it beats both
baselines on all 12 model--task pairs, by 7.2 points over CacheBlend and 3.5
over ProphetKV on average. Its worst relative loss to lossless is 8.3\%, 4.4\%
and 3.3\% at 5\%, 10\% and 15\%, against 44.5\%, 46.2\% and 42.2\% for
CacheBlend and 17.5\%, 14.2\% and 7.7\% for ProphetKV, and from 10\% on it
matches lossless on average. On 6 pairs, \sysname at 5\% already beats both
baselines at 15\%. Unlike them, it never falls below trained PIC: CacheBlend
does on 8 pairs, and ProphetKV, whose gains come primarily from retrieval, does on
every model's QA average.

\autoref{fig:eval-granularity} fixes the budget at 5\% and varies the selector
block size $g$, with one selector trained per $g$; coarser blocks recompute fewer
distinct regions at lower latency and memory cost. In every model's four-task
average, \sysname leads ProphetKV at every block size, by 6.0, 3.4 and 1.7
points at $g=32$ on Qwen3-4B, Qwen3-32B and Gemma4-12B. Its best setting also
beats ProphetKV's best on every model, by up to 2.5 points on DocQA and 9.0 on
RULER MV.

\subsection{Serving Performance}
\label{sec:eval-serving}
\begin{figure}[!t]
\centering
\includegraphics[width=\columnwidth]{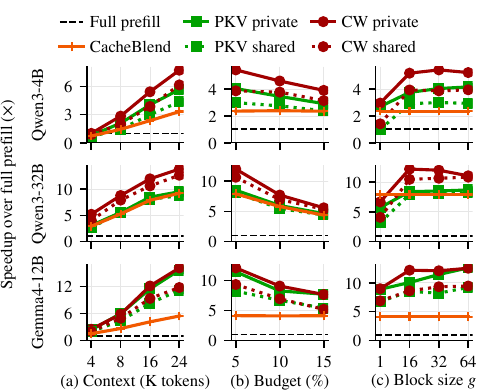}
\caption{Single-request TTFT speedup over SGLang full prefill. We use CacheBlend 
in LMCache and adapt it to support Gemma4-12B. \sysshort/PKV: \sysname/ProphetKV.}
\label{fig:eval-latency}
\end{figure}
\begin{figure}[!t]
\centering
\includegraphics[width=\columnwidth]{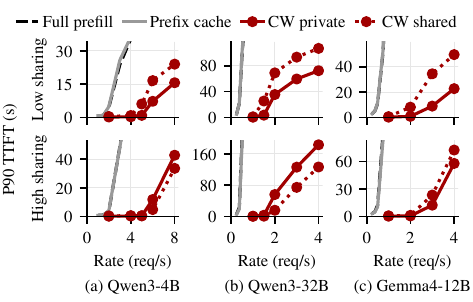}
\caption{P90 TTFT. Under a 5\,s P90, \sysserve sustains 2.5--8$\times$ higher
rates than the baselines. Shared mode trails private under low sharing but is
up to 3.5$\times$ faster under high sharing.}
\label{fig:eval-serving-p90}
\end{figure}
\noindent\textbf{Single request.} \autoref{fig:eval-latency} measures TTFT one
request at a time on one GPU over 30 MuSiQue-derived prompts of around 1K-token
chunks, varying context length, budget and $g$ from a default of 16K, 5\% and
$g=32$. We compare with CacheBlend on LMCache and with our optimized port of
ProphetKV; it and \sysname run in private and shared mode
(\autoref{sec:serving-modes}). Reuse pays off more at longer contexts: at 24K,
\sysname is 7.7--16.1$\times$ faster than full prefill and 1.5--3.0$\times$
faster than CacheBlend in private mode, and 6.2--12.7$\times$ and
1.4--2.2$\times$ in shared mode, which adds sharing overheads
(\autoref{sec:serving-modes}); Gemma4-12B reaches 11.8$\times$ over full prefill.
Recomputing less makes \sysname much faster: a 5\% budget instead of 15\% cuts
TTFT in private mode by 1.4$\times$, 2.1$\times$ and 1.6$\times$ on
Qwen3-4B, Qwen3-32B and Gemma4-12B, as 5\% of 16K is already about 830 tokens, versus a 22--51-token
question. \sysname can exploit this, since at a 5\% budget its four-task
average already matches ProphetKV's at 15\% on every model
(\autoref{fig:eval-accuracy}).
\sysname stays accurate with coarse blocks, losing at most 1.1 points from
$g=1$ to $g=32$ (\autoref{fig:eval-granularity}), which is key to efficiency:
at $g=1$, selections touch 8--10$\times$ more KV blocks, fragmenting attention
and raising TTFT by up to 1.8$\times$ in private and 2.7$\times$ in shared mode.

\noindent\textbf{Under load.} We send MuSiQue-derived requests with Poisson
arrivals, each shaped like the default of \autoref{fig:eval-latency}. Under
\emph{low sharing}, every chunk is unique; under \emph{high sharing}, each group
of 64 consecutive requests shares 62.5\% of its material chunks. All chunks have
their PIC caches prepared in the L2 store before arrivals start. Full prefill
and prefix caching saturate first (\autoref{fig:eval-serving-p90}), since shared
chunks appear at different positions and leave prefix caching little to reuse.
We also calculate SLO goodput, the rate of requests meeting both TTFT $\le5$\,s
and TPOT $\le150$\,ms. \sysserve raises peak goodput over the better baseline by
up to 7.8$\times$ under low and 7.6$\times$ under high sharing, both on
Qwen3-32B.

\subsection{Ablation}
\label{sec:eval-ablation}
\begin{figure}[!t]
\centering
\includegraphics[width=\columnwidth]{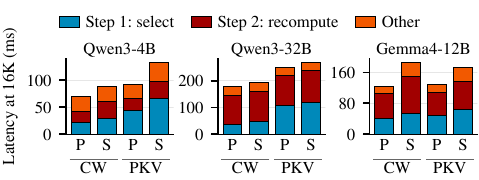}
\caption{TTFT breakdown at \autoref{fig:eval-latency}'s default setting.
P/S: private/shared; Other: cache setup, CoW and serving overhead.}
\label{fig:eval-breakdown}
\end{figure}
\noindent\textbf{TTFT breakdown.} \autoref{fig:eval-breakdown} splits TTFT into
Step 1, the question forward that selects tokens to recompute, and Step 2, the
joint forward of selected tokens and the question (Algorithm~\ref{alg:serving}).
\emph{(1) Selection.} \sysname's lead over ProphetKV comes from Step 1, which is
2.1--2.9$\times$ faster on Qwen3: it scores compact, precomputed block keys
instead of question-to-token attention. The gap narrows on Gemma4, where both
score only the eight full-attention layers. \emph{(2) Recomputation.} Step 2
costs both methods about the same and dominates on Qwen3-32B and Gemma4-12B:
even 5\% (832 tokens) is costly on the larger Qwen3-32B, and Gemma4 also stages
SWA windows. \emph{(3) Sharing.} Shared mode costs more on Gemma4,
due to its 512-dim full-attention heads and SWA handling in
recomputation.

\FloatBarrier

%% file: section/main/discussion.tex
\section{Discussion}
\label{sec:discussion}

\paragraph{SWA storage overhead.} Keeping all SWA KV in L2 adds storage overhead
and limits host cache capacity: for 16K-token chunks on Gemma4-12B, it takes
about $12\times$ the space of a prefix cache that keeps only the trailing
window. To reduce this overhead, bounded replay~\citep{deepseek2026v41flash}
could reconstruct the preceding window for each selected run, with PIC training
adapted to the approximation. Alternatively, cross-layer
sharing~\citep{brandon2024cla} could derive SWA KV from full-attention KV at the
cost of retraining. We leave these directions to future work.

%% file: section/main/conclusion.tex
\section{Conclusion}

PIC training and rule-based recomputation, alone or combined, leave model- and
task-dependent gaps to full-prefill accuracy. \sysname adapts the model to PIC
and trains a model-native selector on its oracles, choosing KV-aligned blocks
without attention scores. \sysserve runs selection and recomputation on the GPU
over private or shared caches. At a 5\% budget, \sysname beats CacheBlend and
ProphetKV by 7.2 and 3.5 points on average; \sysserve cuts TTFT by up to
16.1$\times$ over full prefill and raises goodput by up to 7.8$\times$.